\documentclass[10pt,a4paper,oneside,BCOR=10mm]{article}

\usepackage[onehalfspacing]{setspace}
\usepackage[top=2cm,bottom=3cm,left=2cm,%
right=2cm,bindingoffset=0.0cm]{geometry}
\usepackage[utf8]{inputenc}
\usepackage{amsmath}
\usepackage  [english]{babel}
\usepackage{cite} 
\usepackage{amssymb}
\usepackage{amsmath}
\usepackage{booktabs} 
\usepackage{graphicx} 
\usepackage{makecell}
\renewcommand{\bf}[1]{\textbf{#1}}

\usepackage{enumitem}  
\usepackage[hidelinks, colorlinks]{hyperref}
  \hypersetup{colorlinks,linkcolor=[rgb]{0.2,0.2,0.75},citecolor=[rgb]{0.3,0.5,0.5},urlcolor=[rgb]{0.4,0.4,0.6},breaklinks}
\usepackage{lmodern}
\usepackage{xcolor}

\usepackage{ulem} 

\usepackage{verbatim}
\usepackage{tcolorbox}
\usepackage{color}
\usepackage{siunitx}
\DeclareSIUnit{\rad}{rad}

\usepackage{listings}
\usepackage{authblk}
\usepackage{cleveref}
\usepackage{etoolbox} 

\usepackage[indention=0cm,format=plain,labelfont=bf]{caption}

\newcommand{\q}[1]{``#1''}

\author[1] {F.~Vannini}                           
\author[2] {V.~Bandaru}  
\author[1] {H.~Bergstroem}
\author[1] {N.~Schwarz}
\author[3] {F.~J.~Artola}    
\author[1] {M.~Hoelzl}

\author[1] {G.~Pautasso}
\author[4] {E.~Nardon}
\author[5] {F.~Maviglia}
\author[6] {M.~L.~Richiusa}
\author[7] {E.~Emanuelli}       
\author[8] {the JOREK team} 

\affil[1]{Max Planck Institute for Plasma Physics, Boltzmannstr. 2, 85748 Garching b. M., Germany}                                                     
\affil[2]{Indian Institute of Technology Guwahati, Assam, India}
\affil[3]{ITER Organization, 13067 St. Paul Lez Durance Cedex, France}          
\affil[4]{CEA-IRFM, Saint-Paul-lez-Durance, F-13108, France}   
\affil[5]{EUROfusion Consortium, Boltzmannstr.2, Garching, 85748, Germany}
\affil[6]{UKAEA, Culham Science Centre, Abingdon, Oxon, OX14 3DB, UK}
\affil[7]{NEMO Group, Dipartimento Energia, Politecnico di Torino, Torino, Italy}   
\affil[8]{See author list of M. Hoelzl, G. T. A. Huijsmans, S. J. P. Pamela, M. Becoulet, E. Nardon, F. J. Artola, B. Nkonga et al. Nuclear Fusion 61, 065001 (2021)}    
\title{Runaway electron beam formation, vertical motion, termination and wall loads in EU-DEMO}
\date{}

\begin{document}
\maketitle

\begin{abstract}
Runaway electron loads onto material structures are a major concern for future large tokamaks due to the efficient avalanching at high plasma currents. Here, we perform predictive studies using the JOREK code for a plausible plasma configuration in the European DEMO fusion power plant with focus on a pessimistic scenario in which a multi mega-ampere runaway electron beam is formed. The work first comprises axisymmetric predictions of runaway electron beam formation in a mitigated scenario and of the simultaneous vertical motion of the beam due to loss of position control. The subsequent runaway electron beam termination triggered by a burst of MHD activity during the course of the vertical motion is then simulated in 3D with the runaway electron  fluid self-consistently coupled to the MHD modes. Finally, the resulting deposition pattern of the runaway electrons onto wall structures is calculated with a relativistic test particle approach. This way, the suitability of a possible sacrificial limiter concept for the protection of first wall components is assessed.


\end{abstract}


\newpage
\section{Introduction}\label{sec:introduction}

Disruptions \cite{ITERPhysics,Hender_2007} are major instabilities that can occur in tokamak plasmas terminating the discharge and threatening the integrity of the machine. Since disruptions cause excessive thermal loads onto the plasma-facing components and electromagnetic (EM) forces on the surrounding conductors, their physics must be studied and understood in order to safely operate ITER and other future tokamak devices. Disruptions that follow a vertical displacement of the plasma column from its equilibrium state are known as (hot) vertical displacement events (VDE) \cite{Gruber_1993,LEHNEN201539}. In their presence, the plasma moves towards the wall and part of its current flows into the first wall (halo currents), leading to large Laplace forces on the vacuum vessel and on the in-vessel components. In addition, the reduction of the plasma cross section causes the decrease of $q_\mathrm{95}$. Destabilised 3D MHD instabilities (due to low $q_{95}$) can then cause local heat loads together with toroidal asymmetric halo currents. Of particular concern is the possibility that the rotation frequency of the latter and of their associated EM forces, may resonate with the natural frequencies of the vessel. Major disruptions in which the plasma first loses its thermal confinement, before the vertical instability sets in behave differently in the details, but may cause unacceptable levels of electromagnetic and heat loads as well. For this reason, unmitigated disruptions of any kind will need to be avoided or mitigated in large tokamak devices.
\\
A promising mitigation technique to reduce heat loads and EM forces on the wall is the injection of large quantities of hydrogen/deuterium and impurities when an upcoming disruption is predicted. Impurity injection dissipates most of the thermal energy of the plasma on a millisecond timescale during the thermal quench (TQ) and reduces the total vertical force $F_z$ of the plasma on the wall \cite{Schwarz_2023}. At the end of the TQ, the plasma cools down to temperatures $\sim 10$ eV and the plasma resistivity $\eta$ increases by several orders of magnitude, leading to the decay of the plasma current on a resistive timescale during the current quench (CQ) and to the dissipation of the magnetic energy, which terminates the discharge. However, the appearance of strong self-induced toroidal electric fields during the CQ can accelerate electrons to relativistic velocities. This population of electrons, known as runaway electrons (REs) \cite{Boozer_2018,Breizman_2019}, can form a beam in large fusion machines, carrying a large fraction of the pre-CQ plasma current. Uncontrolled losses of REs to the plasma-facing components can then cause substantial damage and significant melting \cite{Reux_2015}. \\
In view of the construction of the first \q{European DEMOnstration Fusion Power Plant} (EU-DEMO or simply DEMO) \cite{FEDERICI2021112959,Federici2019,Siccinio2020}, methods to avoid, control and suppress RE losses need to be investigated. In addition, as proposed in Ref.~\cite{Barrett_2019}, ``sacrificial'' wall limiters are planned to be implemented. These are discrete plasma-facing components that aim to absorb most of the load to protect the first wall (FW) of the machine in the event of a catastrophic failure.\\ 
The numerical studies carried out here aim to test whether upper limiters (ULs) are capable of protecting the wall from the heat loads of formed RE beams during an upward mitigated hot VDE. To address this question, we use the 3D nonlinear magnetohydrodynamic (MHD) code JOREK \cite{Hoelzl_2021}. Our aim is to contribute to a complete understanding of RE beam formation, vertical motion, termination, and machine protection via a sacrificial limiter by performing predictive numerical simulations that estimate the RE heat loads on the machine FW and to see how these are modified when the presence of the ULs is taken into account. 
\\ We first study upward mitigated VDEs in axisymmetric (2D) simulations, similar to what was done in Ref.~\cite{Schwarz_2023}. There, numerical studies were validated against ASDEX Upgrade and JET experimental data, and predictions for an ITER scenario were presented. However, here we also take into account the presence of the REs and their back-reaction on the background plasma \cite{PhysRevVinodh}. This is done by considering the coupling of the MHD plasma equations with the RE fluid model implemented in JOREK that treats the REs as a separate fluid species. After observing the RE beam formation, we switch to 3D simulations to study the RE beam termination. Finally, in post-processing via relativistic test particle tracing \cite{Sommariva_2018}, RE markers are seeded inside the computational domain. The markers are then evolved in the EM fields of the aforementioned 3D fluid simulations. When lost to the FW or to the ULs (when present), the marker power deposition is  calculated.\\
The paper is structured as follows: \cref{sec:model} presents the model used, while \cref{sec:scenario} describes the scenario considered. In \cref{sec:2D} the RE beam formation is studied in axisymmetric (2D) simulations, during a mitigated upward VDE. In \cref{sec:3D}, the RE beam termination is studied in non-axisymmetric (3D) simulations. In \cref{sec:markers} the RE heat loads on the plasma-facing components are estimated and in \cref{sec:conclusion} the conclusions of this work are drawn.



\section{Model}\label{sec:model}
The results of the simulations that will be presented and analyzed in the next sections have been obtained using the 3D non-linear code JOREK.\\ 
JOREK is based on the right-handed cylindrical coordinate system $(R,Z,\varphi)$, with the toroidal angle $\varphi$ oriented clockwise looking from above the torus.  As described in Ref.~\cite{Hoelzl_2021}, various physical models are available in JOREK. Here, we consider a single fluid representation of the background plasma consisting of ions (\q{i}) and electrons (\q{e}). The background (or thermal, subscript \q{th}) plasma is characterized by an ``MHD'' temperature $T=T_{e}+T_{i}$ assuming $T_e=T_i$. In absence of REs, the total toroidal current density ($j$) corresponds to the total toroidal current density of the thermal plasma ($j_{th}$). Similarly, in absence of impurities the total mass density $\rho$ corresponds to the total thermal plasma mass density $\rho=n_i\,m_i$, being $n_{i}$ the ion number density and $m_{i}$ the ion mass. To reduce computational costs and for the sake of simplicity, we consider a reduced MHD plasma model, obtained by expressing the  magnetic field ($\boldsymbol{B}$) and plasma fluid velocity ($\boldsymbol{v}$) using the following ansatz in the normalized basis $(\hat{\boldsymbol{e}}_R,\hat{\boldsymbol{e}}_Z,\hat{\boldsymbol{e}}_{\varphi})$:
\begin{align}
\boldsymbol{B}=\frac{1}{R}\nabla\psi\times\hat{\boldsymbol{e}}_{\varphi}+\frac{F_0}{R}\hat{\boldsymbol{e}}_\varphi=\frac{1}{R}\partial_Z\psi\,\hat{\boldsymbol{e}}_R-\frac{1}{R}\partial_R\psi\,\hat{\boldsymbol{e}}_Z + \frac{F_0}{R}\hat{\boldsymbol{e}}_{\varphi},\quad
    &
    \quad
    \boldsymbol{v}=-R\nabla u\times \hat{\boldsymbol{e}}_{\varphi}\,.
    \label{EqB}
\end{align}
In \cref{EqB} $\psi$ is the poloidal magnetic flux and $F_0$ is a constant in space and time describing the intensity of the vacuum toroidal magnetic  field $B_{\varphi,0}=F_0/R_0$, being $R_0$ the major radius of the plasma axis. $u$ is the velocity stream function, defined as the ratio between the electric scalar potential $\Phi$ and $F_0$.\\
When included, impurities (subscript \q{imp}) are treated as a separate fluid species, characterized by the total impurity mass density $\rho_{imp}$. They are initialized in the simulation domain through a uniform density source $S_{imp}$ and they are assumed to be in coronal equilibrium for simplicity while a marker based model exists in JOREK as well that traces the full charge state evolution.  The total mass density is then given by the sum of the thermal plasma mass density and that of the impurities $\rho=\rho_{th}+\rho_{imp}$. Impurities are convected together with the thermal plasma. The coupling of the reduced MHD plasma model with the impurity model in use in JOREK, is described in Ref.\cite{Hu_2021}.\\
Also REs, when included, are treated as a separated fluid species with respect to the thermal plasma. In this case, the total toroidal current density is decomposed into the sum of the thermal plasma contribution and that of the REs: $j=j_{th}+j_{RE}$, with the runaway electron toroidal current density $j_{RE}=-e\,c\,n_{RE}$. $n_{RE}$ is the RE number density, $e$ is the elementary charge and $c$ is the speed of light. 
In this work, when included, REs are initialized through a seed $S_{RE}$ due to Tritium decay and Compton scattering. Additionally, the secondary volumetric source of REs is represented by $S_{Avalanche}$ which reproduces the RE seed amplification via large angle knock-on collisions, by means of the Rosenbluth-Putvinski model \cite{Rosenbluth_1997} with additional corrections for partially ionized impurities~\cite{Hesslow2018}. The coupling of the RE fluid model to the MHD equations in JOREK, is described in Ref.~\cite{PhysRevVinodh} while further extensions and benchmarks are contained in Ref.~\cite{VinodhBenchmarks}. 
\\
The variables describing the evolution of our system are then: $(\psi,u,j,\omega,\rho,p,\rho_{imp},n_{RE})$, $\omega$ being the toroidal vorticity and   $p$ the plasma pressure. The (normalized) equations taken into account in this work, governing the evolution of the reduced single-fluid MHD plasma model coupled to the impurities and RE equations, are reported in \cref{sec:appendix}, together with the values of some meaningful used plasma parameters.\\ 
Fixed boundary conditions are considered for all the listed variables (unless stated otherwise), except $\psi$ and $j$. To impose the boundary conditions on $\psi$ and $j$, the coupling with the resistive wall code STARWALL is considered here \cite{Holzl_2012,Artola_2022}. The fully implicit JOREK-STARWALL model takes into account the effects of the conducting structures surrounding the plasma and allows the vertical motion of the latter to be captured. The coupling is obtained via the boundary integral formalism (using Green's functions) at the edge of the JOREK computational domain, so that the simulation domain does not need to be extended beyond the plasma region.\\
\Cref{Eq1,Eq2,Eq3,Eq4,Eq5,Eq6,Eq7,Eq8} written in a weak form, are solved on a 2D Isoparametric Bezier finite element polar grid combined with a Fourier expansion in the toroidal direction. The simple polar grid chosen here for convenience is characterized by a specified number of nodes in the radial $n_{R}$ and poloidal $n_{\theta}$ directions. In \cref{sec:2D} the results of axisymmetric (2D) simulations ($n=0$-only retained, being $n$ the toroidal mode number) will be presented and discussed. The simulations of \cref{sec:3D} focusing on the MHD activity related to the termination, in contrast, are conducted including several  toroidal modes to resolve the 3D dynamics.


\section{Scenario}\label{sec:scenario}
\begin{figure}[!h]
    \centering  
    \includegraphics[width=400pt]{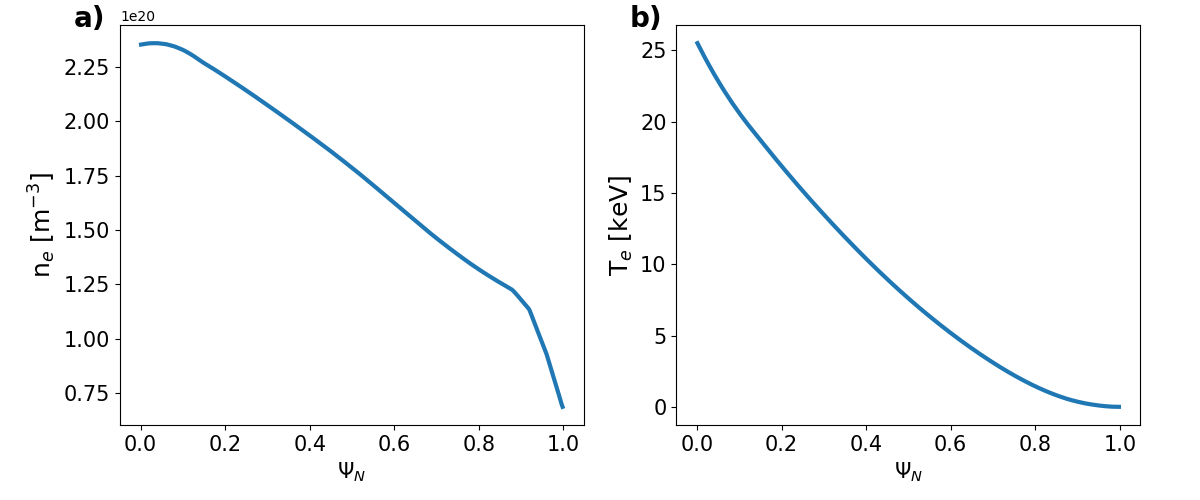}
    \caption{Equilibrium electron density \bf{(a)} and temperature \bf{(b)} profiles.}
    \label{fig:equilibrium_profiles}
\end{figure}

The \q{DEMO Single Null (SN) Variant (2021)} has been selected for our studies. The term \q{variant} refers to a specific machine design characterised by a number of physical and technological constraints. The variant under study has been produced by the system codes PROCESS\cite{KOVARI20143054,KOVARI20169}, while the associated magnetic equilibrium for the start of the flat-top (SOF) has been created by the code CREATE-NL\cite{ALBANESE2015664}. More details about the variant studied can be found in Ref.~\cite{SICCINIO2022113047}.
\begin{figure}[!h]
    \centering  
    \includegraphics[width=250pt]{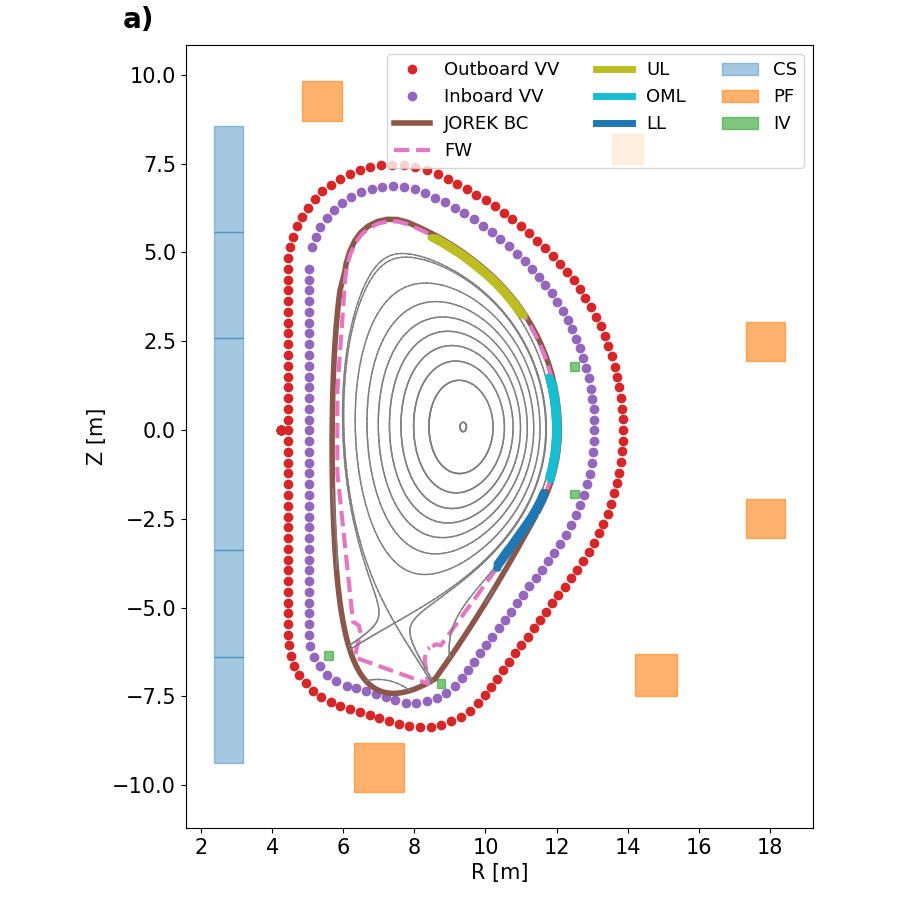}
    \includegraphics[width=180pt]{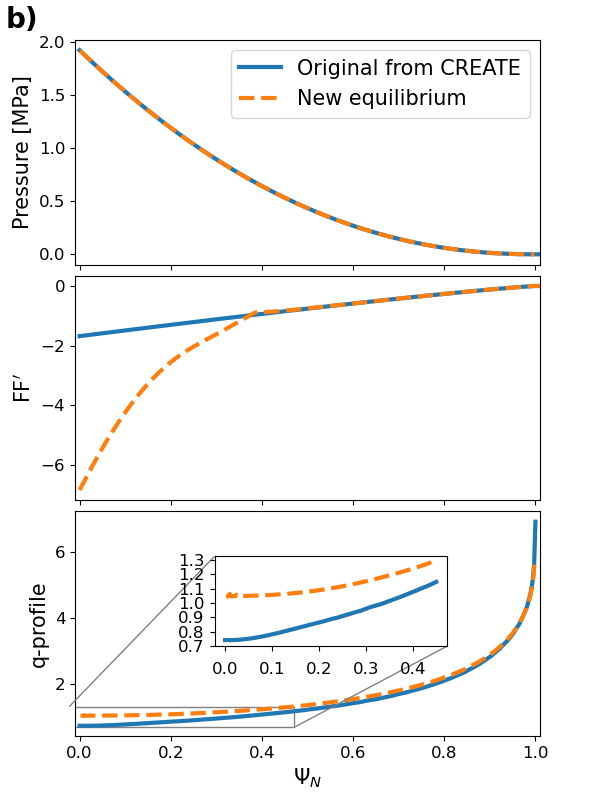}
    \caption{\bf{(a)} Positions of the conducting structures modelled in JOREK-STARWALL according to the specifications of the \q{DEMO Single Null (SN) Variant (2021)}. Inside the FW, the initial magnetic equilibrium of our simulations is shown. The extent of the JOREK plasma boundary conditions (JOREK BC) is also shown (brown solid line). These have been chosen to include the region bounded by the FW and to be as close to it as possible, especially in the upper poloidal plane, that is, the area where the ULs are present. \bf{(b)} From top to bottom: pressure, FF$^\prime$ and corresponding safety factor profiles as a function of the normalised poloidal flux $\psi_N$. The original profiles produced by CREATE-NL (solid blue lines) are compared with those modified by us (dashed orange lines), associated with a new MHD stable equilibrium. The latter are the starting points of our simulations.}
    \label{fig:equilibrium}
\end{figure}
\\
In \cref{fig:equilibrium}\,(a) the positions in the R-Z plane of the conducting structures described in the considered variant and modelled in JOREK-STARWALL are shown. In particular this includes: the components of the central solenoid (CS), the poloidal field coils (PF), the vertical stability coils (IV) and the FW. The outboard and inboard axisymmetric layers of the vacuum vessel (VV) are also shown, each layer having a resistivity of $\eta_w=0.76\mu\Omega\,m$ and a thickness of $d_w=3$ cm. The vacuum vessel layers are modelled in JOREK-STARWALL using the thin-wall approximation. The inboard VV layer is discretized with thin linear triangles, while the outboard VV layer is discretized with toroidal filaments. In \cref{fig:equilibrium}\,(a) the positions of the three different limiter systems are also shown. These are: the upper limiters (ULs), the outer midplane limiters (OMLs) and the lower limiters (LLs). Their positions and shapes have been obtained following the design presented in Ref.~\cite{RICHIUSA2024114329}. The limiters shown appear as segments in the R-Z plane, since their protrusion from the FW is of $\approx 70$ mm. As already discussed in \cref{sec:introduction}, the focus of the present paper is on the capability of the ULs, as presently designed, to shield the FW from the REs. Because of this the adopted JOREK plasma boundary conditions (JOREK BC) have been chosen to contain the FW, being as close as possible to it in particular in the upper part where the ULs are present. In \cref{fig:equilibrium}\,(a) the JOREK BC are indicated by the brown solid line, while the FW is represented by the pink dashed line. The 8 ULs are located every $45^\circ$ in the toroidal direction and placed below the upper port. Each of the 8 units extends for $\Delta\varphi\approx 11.25^\circ$ in the toroidal direction.\\
In \cref{fig:equilibrium}\,(b) the pressure and FF$^\prime$ profiles associated with two different equilibria for the SOF are shown. These, together with the value of the poloidal flux at the JOREK BC, are required by the Grad-Shafranov solver built into JOREK to calculate the initial equilibrium. The safety factor profiles associated with the two different equilibria are shown in the third plot from the top. As the reader can observe, the initial profiles generated by CREATE-NL (solid blue lines) were associated to an MHD unstable equilibrium. Indeed, the values of the corresponding safety factor profile were below $1$ in a large portion of the plane. In order to avoid the growth of unwanted MHD instabilities and for the purpose of our studies, we have modified the original equilibrium as shown in \cref{fig:equilibrium}\,(b) obtaining the orange dashed lines. In particular, we have increased the value of the q-profile to make it larger than $1$ everywhere, by tuning the FF$^\prime$ profile in the core without modifying the original plasma pressure profile. In this way, we produced a new magnetic equilibrium that is MHD stable. As can be observed in \cref{fig:equilibrium}\,(b) in the third plot from the top, the new produced q-profile (orange dashed line) matches the original one for $\psi_N>0.8$, being $\psi_N$ the normalized poloidal magnetic flux. Through this procedure, however, we have slightly modified some plasma parameters as indicated in \cref{Table_modifications}.
\begin{table}[!h]
\begin{center}
\caption{Start of flat-top configuration plasma parameters for the original CREATE equilibrium and for the modified equilibrium. The plasma parameters shown are: total toroidal plasma current ($I_{p}$), internal inductance ($l_i$), poloidal beta ($\beta_{pol}$), magnetic axis positions ($R_{Axis}$ and $Z_{Axis}$). Major ($R_{Geo}$) and minor ($a$) radius, toroidal magnetic field on axis (B$_{tor}$). Value of the safety factor at $\psi_N=0.95$ ($q_{95}$), elongation ($k$) and triangularity ($\delta$).}
\label{Table_modifications}
\begin{tabular}{||c c c c c c c c c c c||} 
 \hline
  & $I_p$ [MA] & $l_i$ & $\beta_{pol}$ & $(R,Z)_{Axis}$ [m] & $R_{Geo}$ [m] & $a$ [m] & B$_{tor}$ [T] & $q_{95}$ & $k$ & $\delta$\\ [0.5ex] 
 \hline\hline
 CREATE-NL & 18.27 & 1.04 & 1.02 & (9.47, 0.06) & 8.94 & 2.88 & 5.7 & 3.51 & 1.77 & 0.44 \\
 \bf{New equilibrium} & 19.73 & 1.08 & 1.01 & (9.37, 0.086) & 8.85 & 2.84 & 5.78 & 3.54 & 1.8 & 0.435 \\ [1ex] 
 \hline
\end{tabular}
\end{center}
\end{table}
\\
The newly produced magnetic equilibrium represents the starting point of our studies and it is characterized by the electron density ($n_e$) and temperature ($T_e$) profiles shown in \cref{fig:equilibrium_profiles}, chosen to match the initial plasma pressure, cf. \cref{fig:equilibrium}~(b).


\section{Mitigated VDE and RE beam formation}\label{sec:2D}
\begin{figure}[!h]
    \centering  
    \includegraphics[width=500pt]{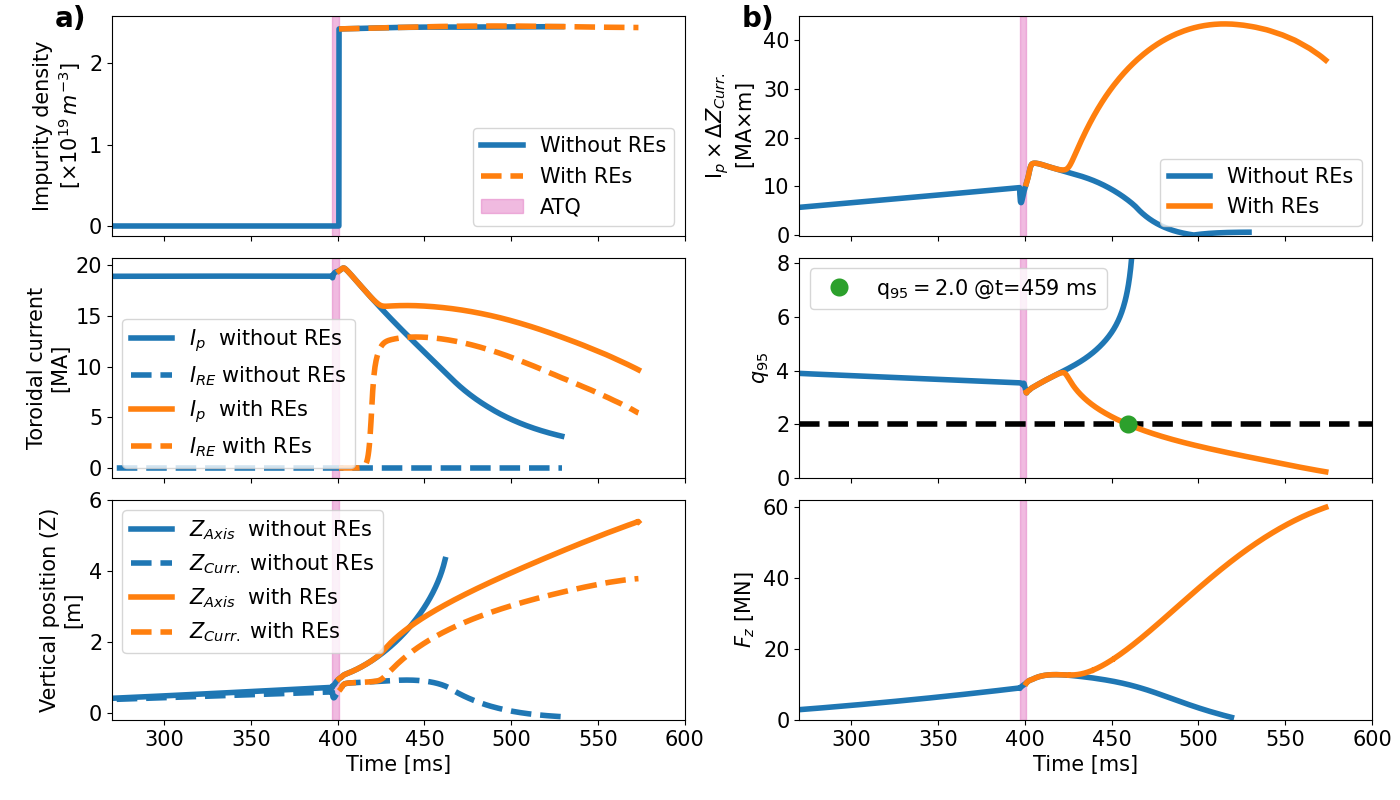}
    \caption{Time traces of some meaningful plasma quantities for two different simulations of a mitigated VDE in DEMO. Impurity (neon-only) concentration of $n_{imp}=2.4\times10^{19}m^{-3}$. In one simulation (blue lines), the presence of the REs is neglected. In the other (orange lines) the RE effects are considered. The pink region labels the width of the ATQ ($\Delta t\approx 4$ms). \bf{(a)} From the top to the bottom: the impurity content (switched on before the ATQ), the total toroidal plasma current $I_{p}$ (solid lines) and the RE toroidal current (dashed lines), the vertical positions of the magnetic axis $Z_{Axis}$ (solid lines) and of the current centroid $Z_{Curr.}$(dashed lines). \bf{(b)} From top to bottom: vertical current moment, q$_{95}$ and vertical force on the wall. }
    \label{fig:MitigatedVDE}
\end{figure}

\begin{figure}[!h]
    \centering  
    \includegraphics[width=260pt]{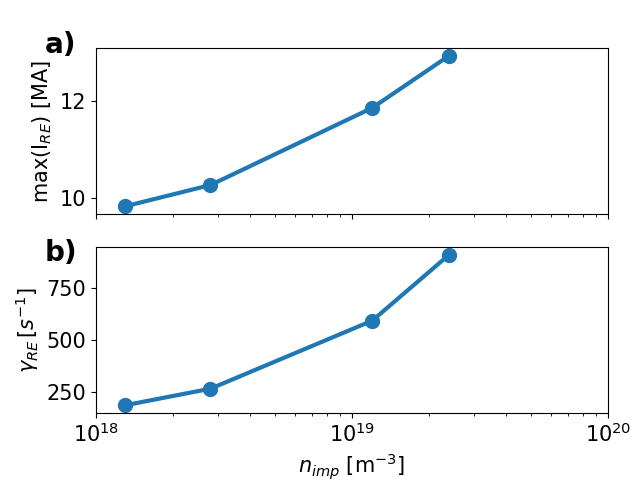}
    \caption{Scan against the amount of impurities injected. \bf{(a)} Maximum value of the toroidal RE current ($I_{RE}$). \bf{(b)} RE beam growth rate ($\gamma_{RE}$).}
    \label{fig:imp_scan}
\end{figure}

As anticipated in \cref{sec:introduction}, we begin by studying an upward hot VDE that would be mitigated after a given displacement by an hypothetical mitigation system. A hot VDE is an initially axisymmetric instability.  Therefore, in this section we discuss simulation results where, for simplicity, only the toroidal mode number $n=0$ is retained for all physical variables of interest. In \cref{fig:MitigatedVDE} the time traces of some meaningful quantities associated with the simulations discussed in the following are shown.\\
We initially discuss the simulation results where the RE presence is not taken into account (blue lines in \cref{fig:MitigatedVDE}). To trigger an upward hot VDE, a current perturbation of the in-vessel coils is added. Once the magnetic axis has vertically moved by approximately $\delta Z_{Axis}\approx 0.7\,m$ from its equilibrium position, the impurities are ramped up inside the full simulation domain until the desired impurity density is reached. Here, neon impurities only are considered. To mimic the effects of impurity injections, the artificial thermal quench (ATQ) is produced. We label this time window with the adjective \q{artificial}, to underline that here the TQ has been produced by manually increasing the thermal plasma diffusivities and by  switching off the Ohmic heating in order to obtain the desired temperature drop ($T_e\approx 10$ eV) at the end of it. Also, the current density profile is flattened via a large hyperresistivity, to reproduce the current redistribution and associated current spike typically observed in tokamak disruptions. The temporal extent of the ATQ in \cref{fig:MitigatedVDE} is denoted by the pink temporal window. At the end of the ATQ, all the plasma parameters are switched back to those of the pre-ATQ and the Ohmic heating is switched on again. The CQ phase follows after the ATQ, because of the low temperature determined by the balance of Ohmic heating and radiation, cf. the second plot from the top in \cref{fig:MitigatedVDE}\,(a) where the total toroidal plasma current ($I_{p}$) decay is observed. Note that the toroidal RE current (blue dashed line) is here identically zero since we performed this initial simulation without including the RE effects. In the third plot from the top of \cref{fig:MitigatedVDE}\,(a), the vertical position of the magnetic axis is shown ($Z_{Axis}$, blue solid line). The magnetic axis accelerates towards the wall and the core area shrinks, with the plasma minor radius $a$ decreasing. The faster plasma current decay with respect to the decrease of the square of the plasma minor radius, determines the increase in the value of $q_{95}$, given that: $q_{95}\propto a^2/I_{p}$. On the other hand, the vertical force on the wall is limited to a maximum value (in this case of $F_z\approx 13$ MN). These results are in agreement with the findings  presented in Ref.~\cite{Miyamoto_2011,Schwarz_2023}. As described there, the vertical force on the wall is proportional to the change in the vertical current moment\cite{Gruber_1993}:
\begin{equation}
    F_z\propto I_{p}\times\Delta Z_{Curr.},\quad\text{being}\quad Z_{Curr.}=\frac{\int j\,Z\,dZ\,dR}{I_{p}}\,.
    \label{Eq:MitigatedVDE}
\end{equation}
As we can observe by looking at the third plot from the top of \cref{fig:MitigatedVDE}\,(a), while the movement of the magnetic axis accelerates towards the wall, the vertical position of current centroid ($Z_{Curr.}$, blue dashed line) remains closer to the midplane as a large fraction of plasma current is induced outside the last closed magnetic flux surface (LCMFS) in the halo region (halo currents). Because of this stagnation of the current centroid, the change in the vertical current moment is limited to a maximum value and hence, also $F_z$. 
\\
We can now ask ourselves, how the presence of REs modifies the results previously described. To address this question, we repeat the previous simulation, but initializing a RE seed at the end of the ATQ. This seed intends to represent the REs generated via the mechanism of Compton scattering and Tritium decay (Dreicer and hot-tail are not considered hereby). The results of this new simulation, shown in \cref{fig:MitigatedVDE} by the orange lines, are in agreement with the studies presented in Ref.~\cite{Vinodh2D} for ITER. In the second plot from the top of \cref{fig:MitigatedVDE}\,(b), we observe the growth of the toroidal RE current (dashed line, $I_{RE}$) due to the conversion of the thermal plasma current by means of the avalanche mechanism, till it reaches the maximum value of $I_{RE}\approx 13$ MA. Note that the difference between the total toroidal plasma current ($I_{p}$, orange solid line) and the toroidal RE current, corresponds to the toroidal halo current. The so formed RE beam slows down the plasma current decay rate, since REs do not decay on a resistive time scale. Unlike the case without REs, the value of $q_{95}$ decays, cf. in \cref{fig:MitigatedVDE}\,(b) the second plot from the top. This happens, because in this case the plasma shrinks faster than the current value decreases ($q_{95}\propto a^2/I_p$). Additionally, the current centroid does not stagnate anymore since a large portion of the current remains inside the LCMFS. Consequently, the vertical current moment and the vertical force on the wall keep increasing. Our simulations already show the saturation of the vertical current moment for this case and a modification in the flattening of the temporal trace of $F_{z}$ that could indicate the achievement of its maximum value, before it decays. However, given the purpose of this paper and given the not unrealistic drop of $q_{95}$ to extremely low values (since the 2D simulation setup used here does not allow for MHD instabilities), we will not study this case further. For completeness, we present in \cref{fig:imp_scan}, a scan in the total amount of injected impurities where the modification of maximum value of the RE current and of the growth rate of the RE beam ($\gamma_{RE}$) are shown. 


\section{RE beam termination}\label{sec:3D}
In the present section, we aim to study the termination of the formed RE beam. To this extent, we follow closely the approach introduced in Ref.~\cite{Vinodh3D} in the context of ITER simulations. \\ 
We consider here the case at highest impurity concentration of n$_{imp}=2.4\times 10^{19}\,m^{-3}$, whose time traces have been shown in \cref{fig:MitigatedVDE}. This choice is motivated by the purpose of this paper that is to study the impact of the REs on the ULs to test whether their design can effectively shield the FW. To this extent, we take into account the worst case scenario obtained, characterized by the highest RE beam produced. The 2D simulation discussed in the previous section represents an \q{intermediate} step toward the study of the RE beam termination. In fact, the strategy here adopted, is to follow the 2D simulation with initialized REs (orange lines in \cref{fig:MitigatedVDE}) and then to choose a time where the value of q$_{95}$ is close to 2. At this selected time, the MHD simulations are restarted in 3D, i.e., retaining higher (than $0$) toroidal harmonics for all the physical variables of interest. Three times have been considered to restart the simulations. Each of these corresponds to a different value of $q_{95}$, namely: $2.7$, $2.29$ and $2.05$ (we remind that $q_{95}$ is decreasing in time).  For all these selected cases we have observed, similarly to the findings for ITER presented in Ref.~\cite{Vinodh3D}, the plasma to be MHD unstable with the different modes initialized growing exponentially in time until saturation is reached. 
\begin{figure}[!h]
    \centering  
    \includegraphics[width=400pt]{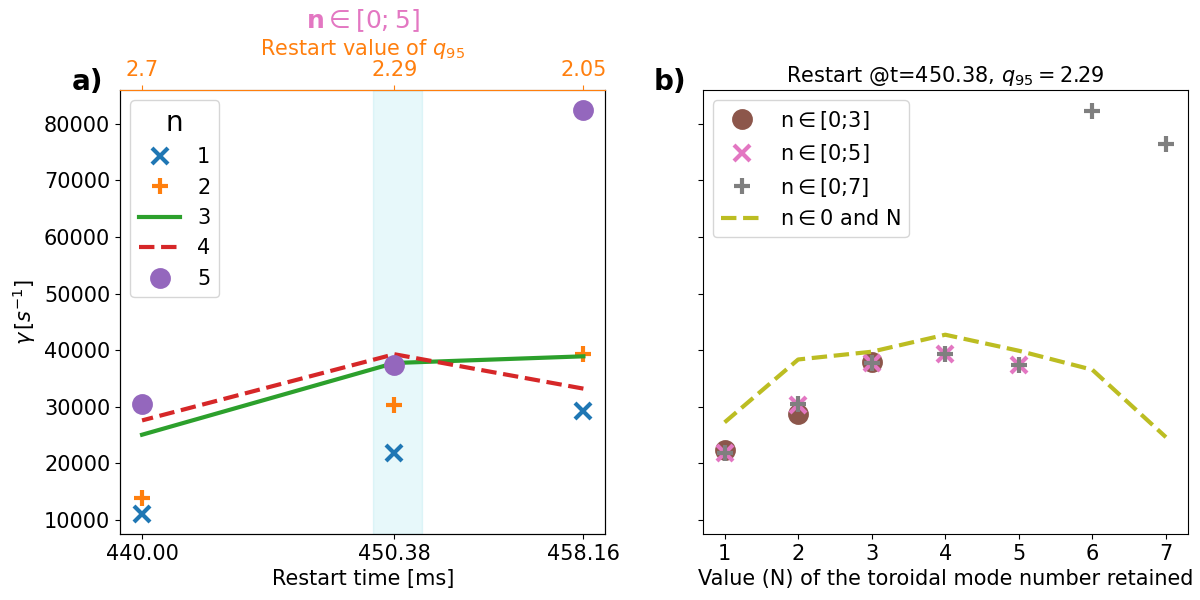}
    \caption{
    \bf{(a)} Poloidal magnetic energy growth rates $\gamma$ associated with instabilities with the toroidal mode number $n$ given in the legend. The results shown correspond to three different simulations, all with $n\in[0;5]$. The 3D restarts were performed at three different times of the 2D simulation with REs in \cref{fig:MitigatedVDE}. These times correspond to different values of q$_{95}$ as indicated on the x-axis. \bf{(b)} Poloidal magnetic energy growth rates of instabilities measured in 3D simulations restarted when $q_{95}=2.29$ of the 2D simulation with REs in \cref{fig:MitigatedVDE}. As indicated in the legend, different toroidal mode numbers were kept in the simulations. The yellow dashed line shows the results of seven different simulations, in each of which only two toroidal mode numbers were retained: $0$ and the value $N$ indicated on the x-axis.
    }
    \label{fig:3Dscan}
\end{figure}
\\
In \cref{fig:3Dscan}\,(a) the growth rates $\gamma$ of the magnetic energies of the instabilities, measured in the linear phase of simulations where $n\in[0;5]$, are shown. There, the 3D restart were performed, as indicated on the x-axis, at the three different times considered, each of which corresponds to one of the chosen values of $q_{95}$.  The growth rates depend on the chosen restart time. Since here we are more interested in the RE beam termination, which occurs after the saturation of the MHD instabilities and during the nonlinear phase, and given that the time traces of the relevant plasma physical quantities exhibit a qualitatively similar behaviour independently from the chosen restart time, we will focus our attention on 3D simulations restarted when $q_{95}=2.29$ ($t=450.38~ms$).  As already discussed in Ref.~\cite{Vinodh3D}, the RE beam can become unstable a long time before the value of $q_{95}=2.0$. Because of this, we do not present here the results produced in the 3D simulations restarted when $q_{95}=2.05$, as this value is too close to the threshold value. In principle, we should further investigate the 3D simulations restarted when $q_{95}=2.7$ or at previous times. However, non-axisymmetric simulations conducted over a long temporal range become challenging and computationally expensive. Because of this, we focus our attention on the 3D simulation restarted at an intermediate time, that corresponds to the value of $q_{95}=2.29$. We also emphasize here, that the increased computational costs have required the use of the numerical scheme described in Ref.~\cite{Holod_2021}, that has improved the simulation time performance. Without it, the simulation results here presented would not have been possible.\\
In \cref{fig:3Dscan}\,(b) the growth rates of the MHD instabilities initialized in simulations restarted when the selected value of $q_{95}=2.29$ has been reached, are shown. There, as indicated in the legend, the results of three different simulations are shown. In each of these, the maximum value, up to which all the toroidal mode numbers have been retained in the simulations, has been varied: $n\in[0;N]$ with $N=\{3,5,7\}$. These growth rates are compared with the results expressed by the dashed line which shows the results of several two-toroidal mode simulations where $n=0$-only and $n=N$-only have been retained (without retaining the instabilities associated to intermediate toroidal mode numbers: $0<n<N$). The retained value of $N$ is indicated on the x-axis. This scan shows us that in the present case $n=4$ is the fastest growing mode and that we should include in our simulations at least all the toroidal mode numbers up to this number to correctly reproduce the linear dynamics. Additionally, the modes $n=6,7$ appears to be nonlinearly driven when included.
\begin{figure}[!h]
    \centering  
    \includegraphics[width=450pt]{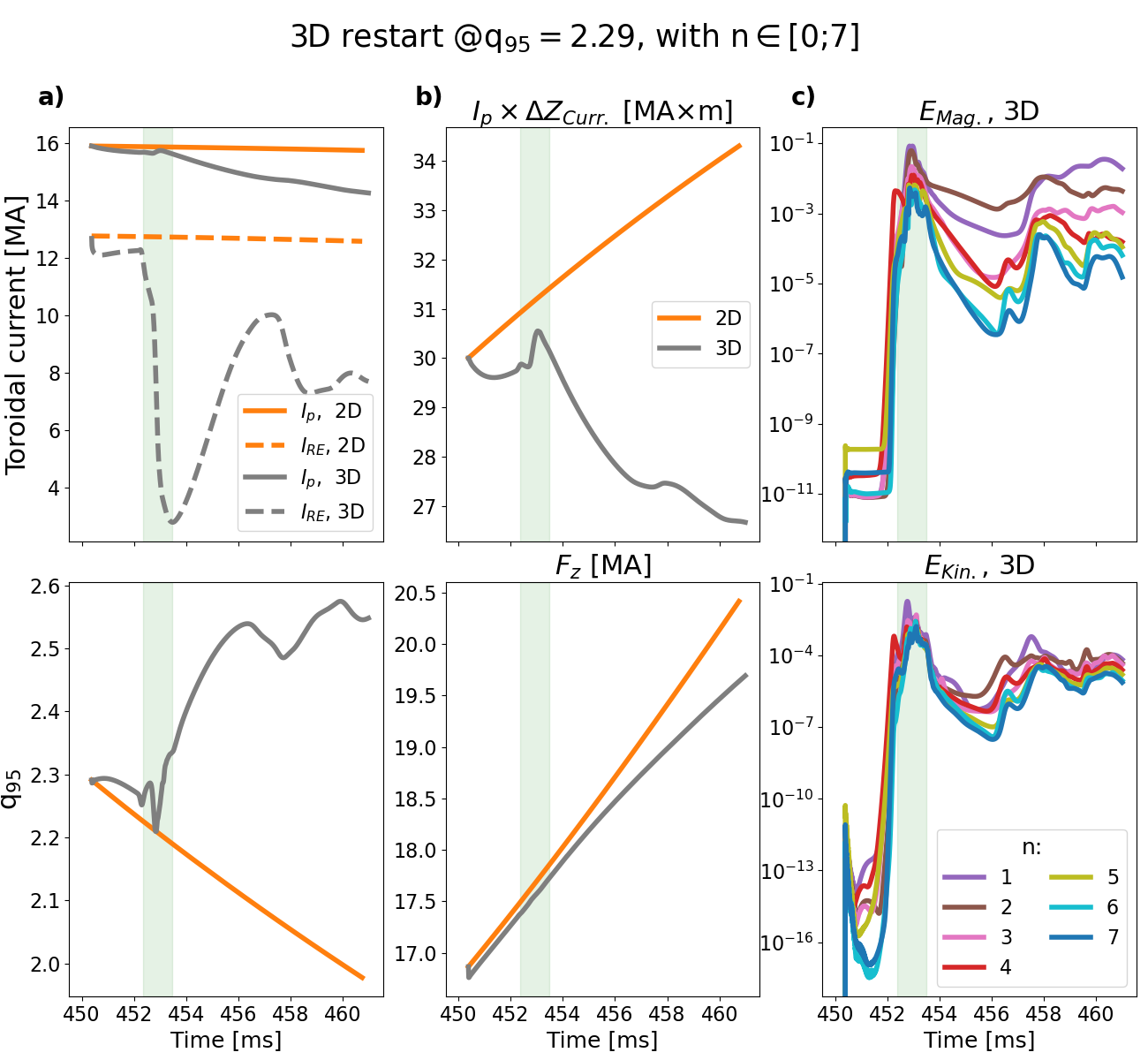}
    \caption{Portion of the axisymmetric simulation (2D) time traces shown in \cref{fig:MitigatedVDE}, compared with those obtained from a 3D restart at $q_{95}=2.29$ with higher toroidal mode numbers retained: $n\in[0;7]$. The green temporal window marks the temporal domain where the RE current decay is observed. The quantities plotted are described from the top to the bottom. \bf{(a)} Toroidal plasma current (solid lines) and RE current (dashed lines), temporal evolution of edge safety factor.  \bf{(b)} Vertical current moment, vertical force on the wall. \bf{(c)} Temporal evolution of poloidal magnetic ($E_{mag}$) and kinetic ($E_{kin}$) energies associated to the initialized instabilities.}
    \label{fig:3Dsims_ntor15}
\end{figure}
\\
In \cref{fig:3Dsims_ntor15} we show again time traces of the 2D simulation (orange lines) already presented in \cref{sec:2D}, zooming in a selected temporal range. These time traces are compared with those (gray lines) obtained in a 3D simulation where $n\in[0;7]$. This corresponds to the simulation with highest toroidal mode numbers retained in this work. In all the plots presented in \cref{fig:3Dsims_ntor15}, six times have been selected, as indicated by the vertical dashed lines. The initial three have been taken at the beginning, in the middle and at the end, respectively, of the RE toroidal current decay phase that is marked by the green temporal domain. At these selected times, the Fourier decomposition of the poloidal magnetic flux has been calculated (cf. \cref{fig:Flux}), together with the Poincar\'e plots (cf. \cref{fig:PP}~(a)) and the safety factor and plasma current profiles (cf. \cref{fig:PP}~(b)). Also, in \cref{fig:Density} the RE density ($n_{RE}$) is shown in the upper part of the tokamak poloidal cross section, on top of which the Poincar\'e plots are shown. All together, \cref{fig:3Dsims_ntor15,fig:Flux,fig:PP,fig:Density} provide us a clear picture of the physics involved here. \\
In \cref{fig:3Dsims_ntor15}\,(c) the poloidal magnetic ($E_{mag}$) and kinetic ($E_{kin}$) energies associated to the non axisymmetric instabilities, are shown. The fastest growing and first saturating mode here is $n=4$. $n=1$ and $n=2$, on the other hand, reach higher saturation levels and remain the dominant instabilities throughout the entire nonlinear phase. We begin by discussing the physics observed in the green temporal range in \cref{fig:3Dsims_ntor15}, referring also to the plots in \cref{fig:Flux,fig:PP,fig:Density} calculated at the times inside this temporal range: $452.36$, $452.84$ and $453.48\,ms$. At the very beginning of this temporal range, $(m,n)=(8,4)$ and $(m,n)=(9,4)$ are the dominant mode structures, $m$ being the poloidal mode number. These are localized close to the edge. Later $(2,1)$ reaches a higher saturation level, resulting in the dominant instability inside the selected time window and remaining localized in the middle of the radial plane at $\rho_N=0.6$. $\rho_N$ being the square root of the normalized poloidal magnetic flux, $\rho_N=\sqrt{\psi_N}$. The development of the MHD instabilities is held responsible for the stochastization of the magnetic field lines. The stochastic region covers initially the outer part of the radial plane and later grows radially inward with the creation of magnetic islands and with the reduction of the closed flux-surface region. Because of the fast RE parallel transport and of the increased stochastic region, REs are progressively lost, while a remaining portion remains concentrated in the core, as it can be observed in \cref{fig:Density}~(a). This causes the drop in the RE current (\q{RE beam termination}) observed in \cref{fig:3Dsims_ntor15}~(a) in the first plot from the top (gray dashed line) with the RE current passing, in the present scenario, in a time window of $\Delta t=1.12\,ms$ from a maximum value of $12.3\,MA$ to a minimum of $2.8\,MA$. The drop in the RE current affects the vertical current moment and slightly reduces the vertical force on the wall.\\
\begin{figure}[!h]
    \centering  
    \includegraphics[width=500pt]{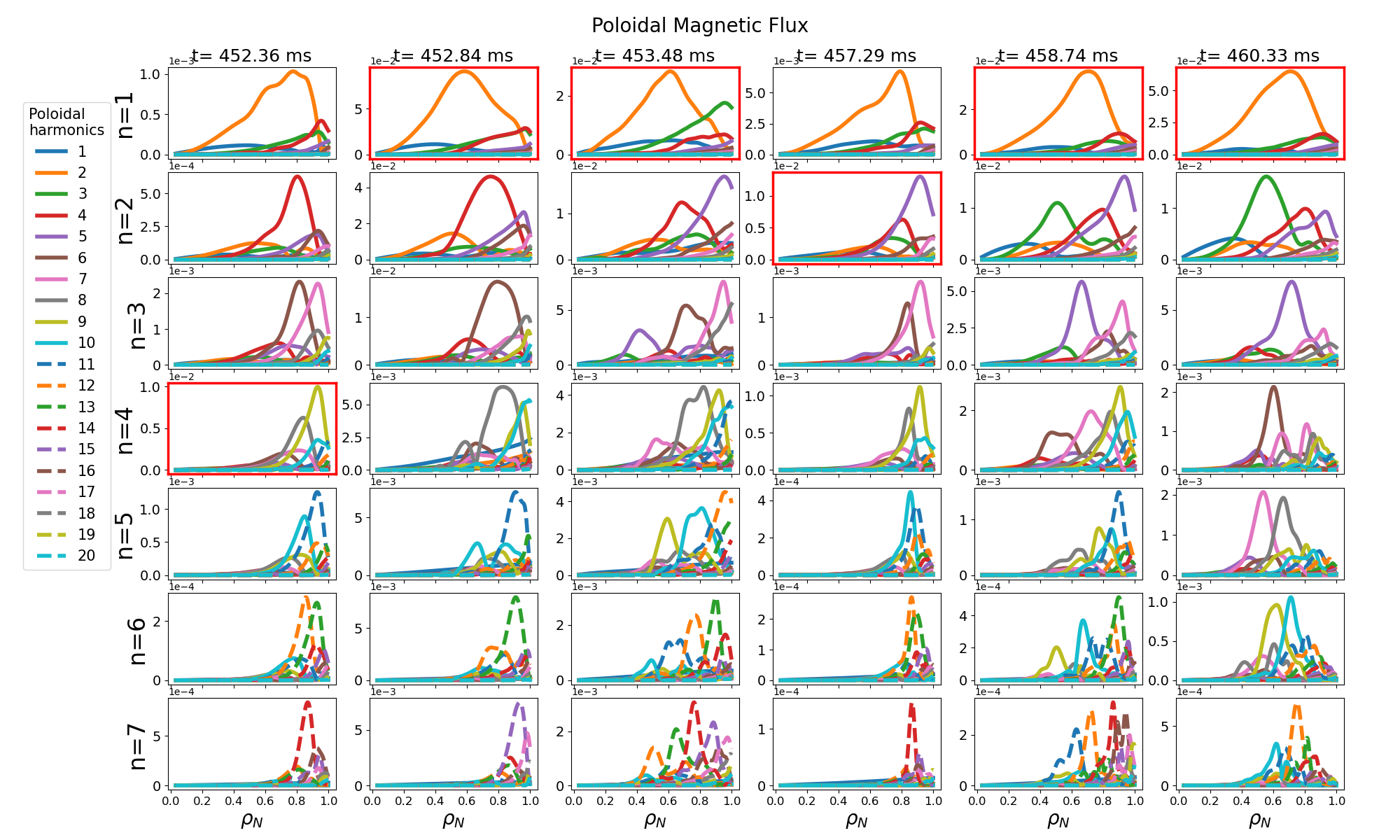}
    \caption{Fourier decomposition of the poloidal magnetic flux at different selected times. For every toroidal mode number retained in the simulation, the mode structure of the dominant poloidal harmonics $m$ is shown. The times at which the mode structure is shown, correspond to those indicated in \cref{fig:3Dsims_ntor15} by the vertical dashed black lines. At each time considered, the dominant mode structure (higher in amplitude), has been marked using red corners.}
    \label{fig:Flux}
\end{figure}
\begin{figure}[!h]
    \centering  
    \includegraphics[width=250pt]{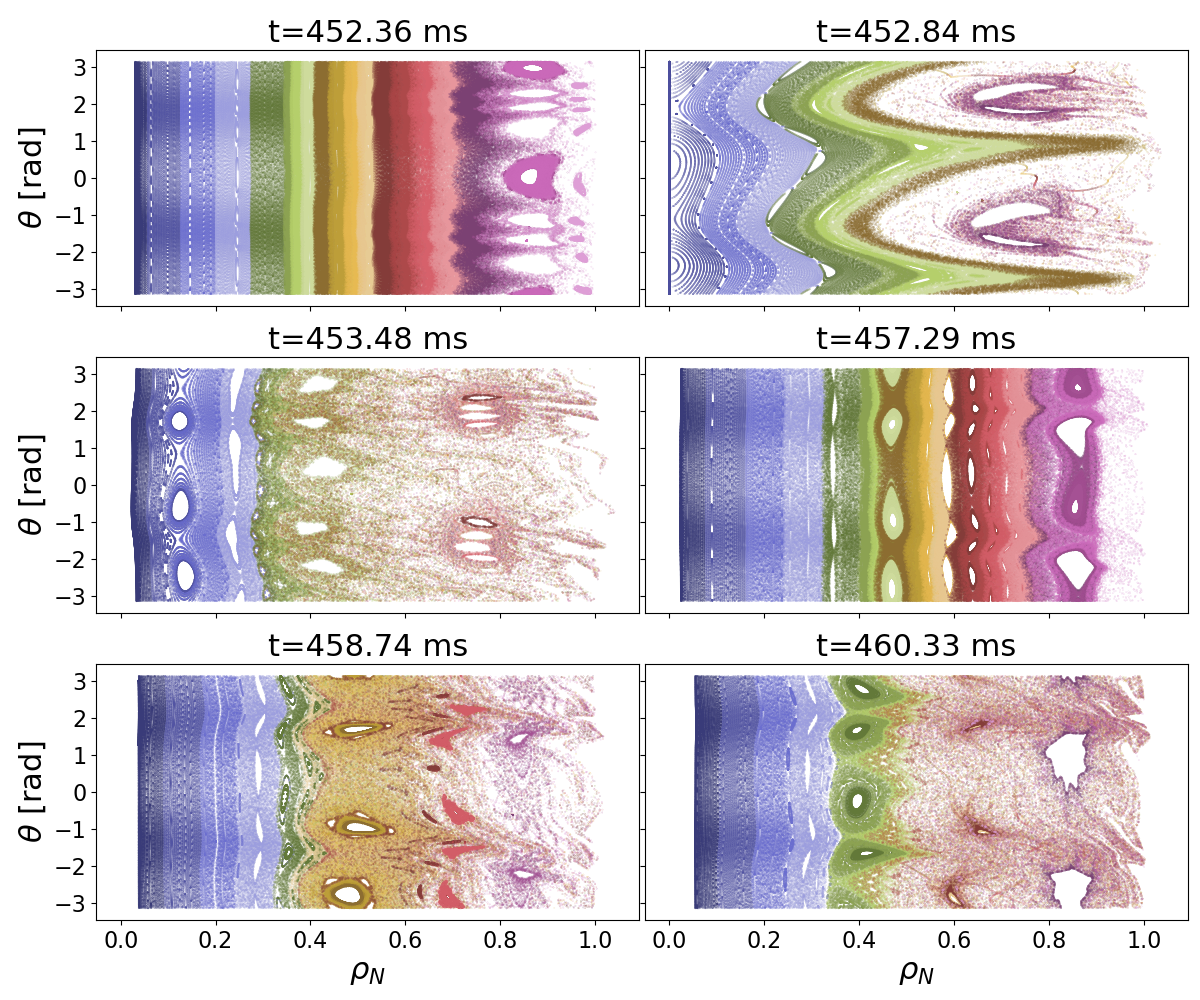}
    \includegraphics[width=230pt]{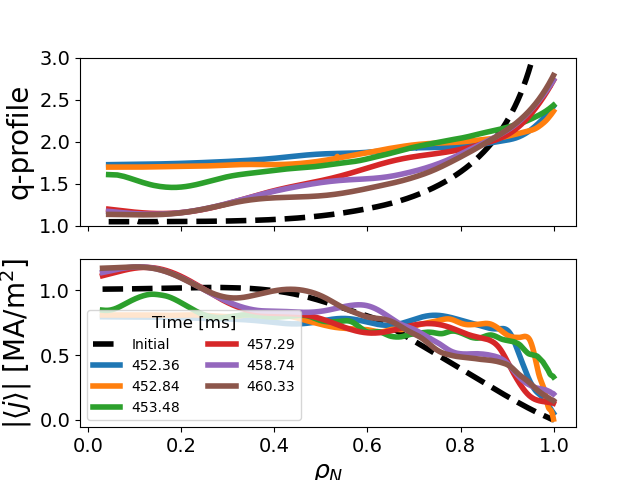}
    \caption{\bf{(a)} Poincar\'e plots at the times at which the Fourier decomposition shown in \cref{fig:Flux} has been performed. \bf{(b)} Shape of the safety factor and flux surface averaged toroidal plasma current at the times at which the Poincar\'e plots have been shown.}
    \label{fig:PP}
\end{figure}
 We observe that, at the end of the RE current decay, a RE seed remains concentrated into the core. The remaining RE current is observed later to re-avalanche, growing again into a beam with maximum current of $10\,MA$. The formed beam slightly decays again after a new burst of MHD activity and so on. This RE current reformation and subsequent decay is observed inside the deep nonlinear phase of the time evolution of the established MHD instabilities. With reference to \cref{fig:3Dsims_ntor15}, we are interested in the temporal dynamics on the right side of the green temporal window. Inside this temporal phase, we have selected three times at which the Poincaré plots and mode structures have been presented. In particular, the first selected time ($t=457.29\,ms$) has been taken when the maximum value in the RE current re-avalanche has been reached. We observe, at this time, the dominant MHD instability that has slightly moved outward, presenting a peak at $\rho_N=0.85$. In addition, the closed flux-surface region has increased again. At the following times, we observe the closed flux-surface region newly decreasing at the expense of the stochastic region. This is accompanied by a partial drop in the RE density and current. In future dedicated works, further studies will be conducted with a particular focus on the physics involved inside this shortly described nonlinear phase of the MHD instabilities. This will be done, also, by presenting simulations with a higher retained number of toroidal modes with respect to those considered here. Given the purpose of the present paper, we do not further investigate the physics involved into the nonlinear phase and we limit our attention to the temporal range where the first and main RE current drop is observed (green temporal region in \cref{fig:3Dsims_ntor15}).

\begin{figure}[!h]
    \centering  
    \includegraphics[width=500pt]{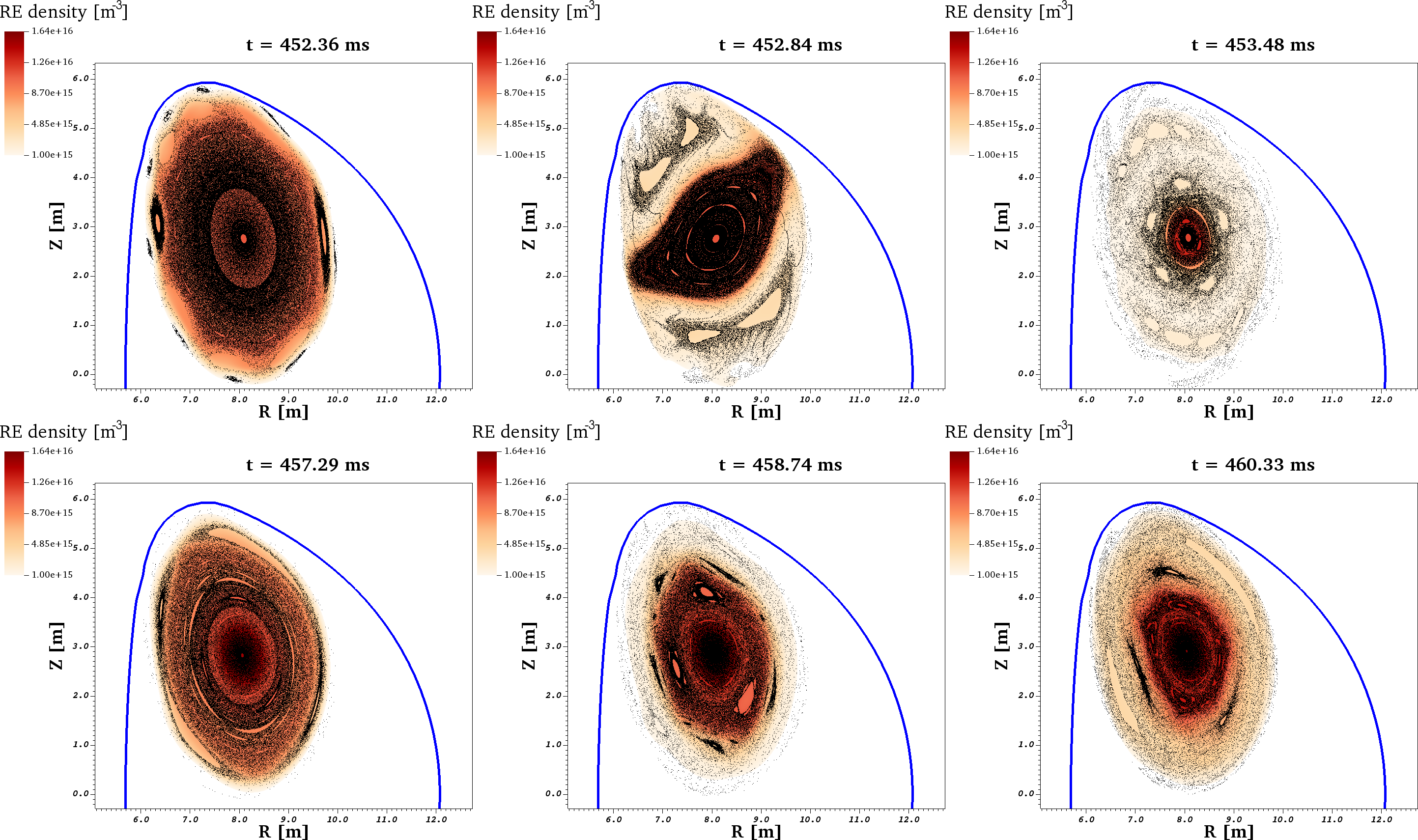}
    \caption{In the upper portion of the poloidal cross section, where the plasma column is located, the RE density is shown at the selected times considered in \cref{fig:Flux}. On top of these, the Poincaré plots are shown. The blue continuous line indicates the FW position.}
    \label{fig:Density}
\end{figure}


\section{Prediction of the RE power deposition on FW and ULs}\label{sec:markers}
\begin{figure}[!h]
    \centering  
    \includegraphics[width=0.8\textwidth]{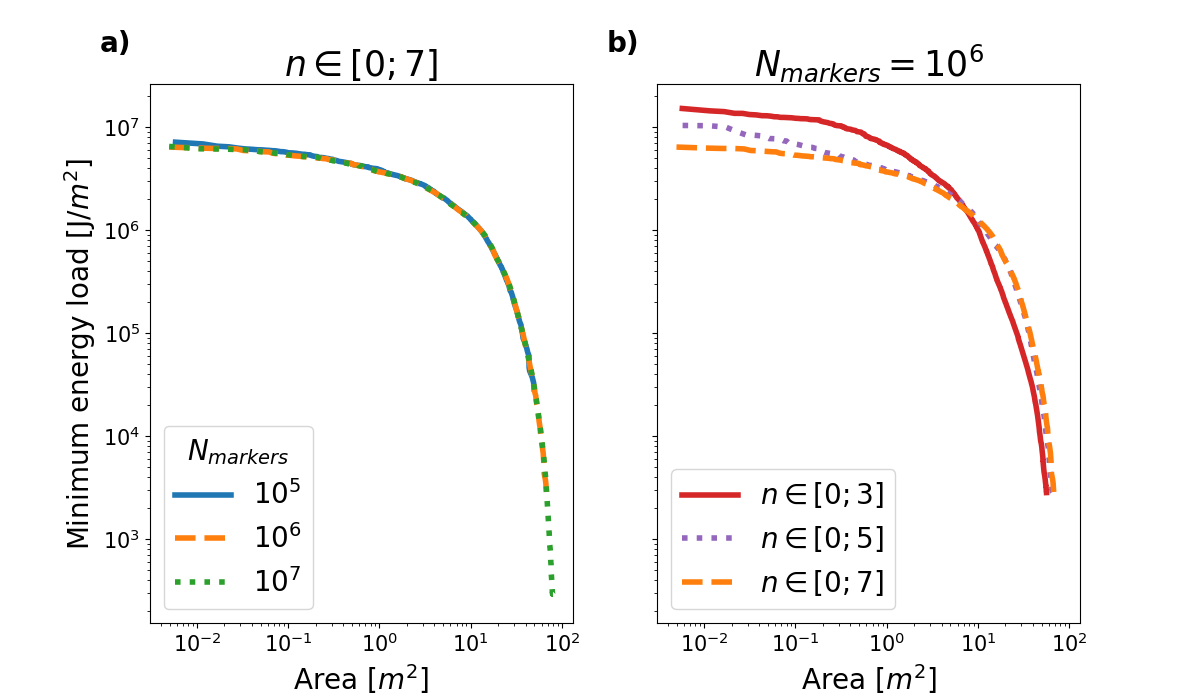}
    \caption{Minimum energy load on the AFW in plasma-wall configurations where the ULs are absent (AFW-only). \textbf{a)} Scan in the number of markers initialized, for simulations with fixed toroidal mode number retained: $n\in[0;7]$. \textbf{b)} Scan in the number of retained toroidal mode numbers in the 3D restarts at $q_{95}=2.29$.  Initialized markers: $N_{markers}=10^6$.}
    \label{fig:ScanEload}
\end{figure}

This section represents the core of the present paper. Here, the power deposited by the REs on FW and ULs is estimated. To do so, we consider the MHD simulation already presented in \cref{sec:3D}, obtained with a 3D restart at $q_{95}=2.29$ where $n\in[0;7]$ have been retained. In post-processing, the tracing marker module available in JOREK\cite{VanVugt,Sommariva_2018} has been used to evolve the RE markers. Similarly to what was done in Ref.~\cite{Hannes2024}, a test-particle approach is used. RE markers, that is super-particles representing a portion of physical particles in phase-space, are evolved using the fields evolution history previously calculated in the corresponding MHD simulation. This implies that the RE markers are not self-consistently coupled to the thermal plasma using a full-$f$ particle-in-cell model like in Ref.~\cite{Hannes2023}. 
In this paper we consider an axisymmetric FW (AFW), generated using a triangular mesh as in Ref.~\cite{Hannes2024}. As mentioned in \cref{sec:scenario}, the AFW is located inside the JOREK BC, cf. \cref{fig:equilibrium}~(a). In fact the JOREK BC were chosen to contain the FW by staying as close to it as possible, especially in the upper part of the tokamak.\\
The RE markers have been initialized at $t=452.25\,ms$, that is the time when the RE current reaches a maximum of $I_{RE}=12.3\,MA$ before it begins its drop during the termination phase, cf.~\cref{fig:3Dsims_ntor15}. The markers are initialized inside the computational domain (JOREK BC) with positions in space that are sampled from the RE density number $n_{RE}$ at the selected time. The same weight, representing the number of physical particles, is associated to all the initialized markers, such that the sum of all the markers weights equals the number of physical RE particles, that is $N_{RE}=1.24\cdot 10^{19}$. The kinetic energy of all the RE particles equals the total magnetic energy ($W_{tot}$) channeled to the REs from the poloidal magnetic field \cite{Vinodh2D}:
\begin{equation}
    W_{tot}=\int j_{RE,\parallel}(E_{\parallel}-E_{c}^{eff})\,dV\,dt\,,
\end{equation}
$j_{RE,\parallel}$ being the parallel RE current density, $E_{\parallel}$ the parallel electric field and $E_{c}^{eff}$ the effective critical electric field \cite{Hesslow_2018}. Here $W_{tot}\approx 35\,MJ$. Since by using fluid simulations we do not have any information about the RE distribution in velocity space, we consider for simplicity the total kinetic energy to be equally distributed among all the particles, so that  every RE particle has an initial energy of $\mathcal{E}=17.6\,MeV$ and pitch angle of $\xi=v_{\parallel}/v=-0.99$. The initialized markers are then traced by solving the guiding center equations \cite{Tao2007} via a fourth order Runge Kutta method (RK4). The markers are evolved until they collide with the AFW (contained inside the JOREK BC) or the minimum value of the RE current at the end of the termination phase has been reached (cf. the green temporal window in \cref{fig:3Dsims_ntor15}). The markers temporal evolution is stopped at $t=453.59\,ms$, before the beginning of the RE beam reformation. In case the markers collide with the wall, they are considered lost and their energy is deposited on the plasma facing components.
\begin{figure}[!h]
    \centering  
    \includegraphics[width=1.1\textwidth]{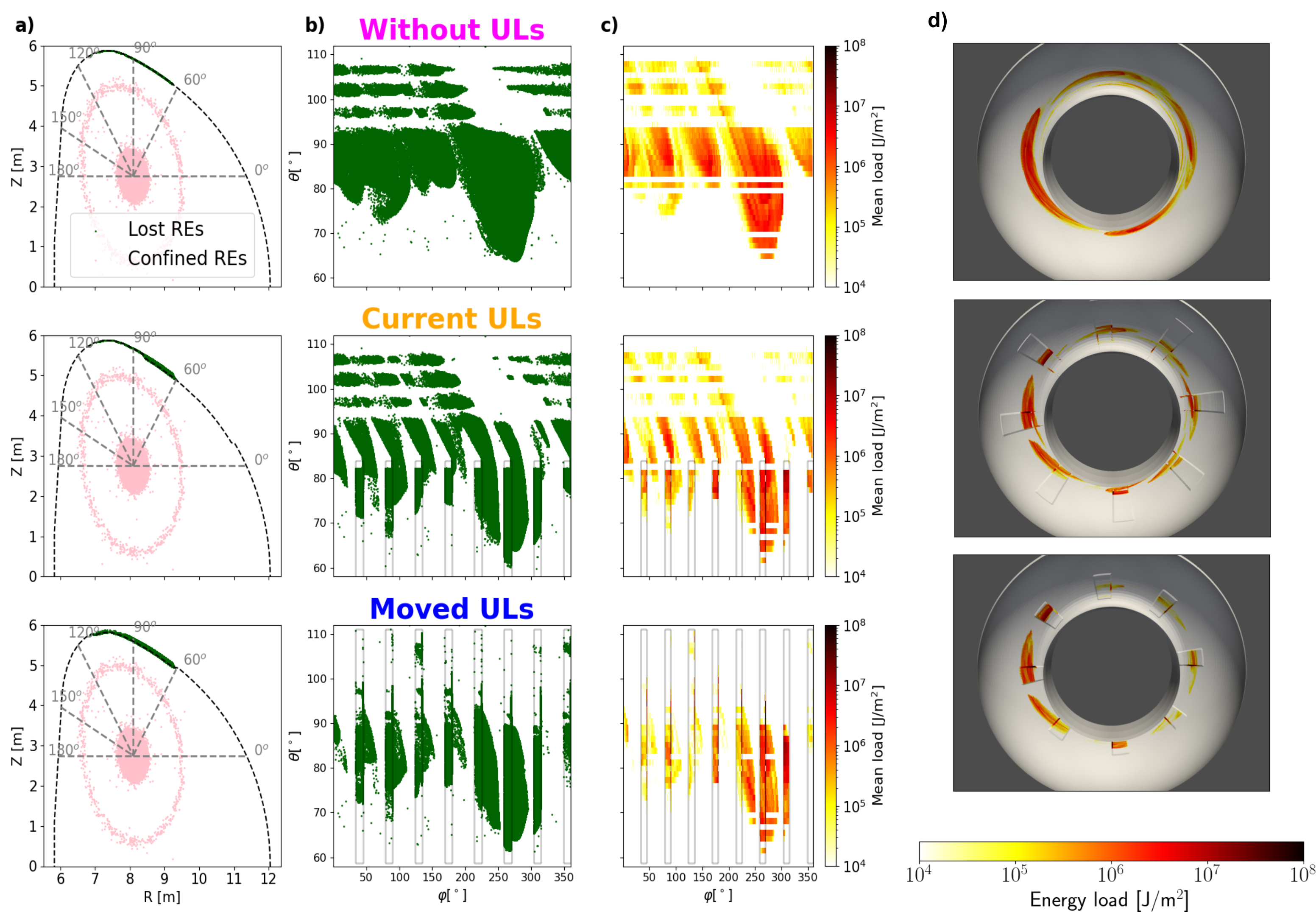}
    \caption{Results at the end of the markers time evolution during the termination phase in \cref{fig:3Dsims_ntor15}: $t\in[452.25;453.59]\,ms$. Here, the initialized number of markers is $N_{markers}=10^6$. Different wall configurations have been considered: in the first line only the presence of the AFW is taken into account, while in the second line also the UL positions as conceived by the DEMO team, are considered. In the third line, the ULs have been moved \q{ad-hoc} to the positions where the majority of the REs hit the wall. \textbf{a)} Confined (pink) and lost (green) markers in the upper part of the R-Z plane. The black dashed lines represent the positions of the plasma-facing components. \textbf{b)} Markers lost to the plasma-facing components, plotted against the toroidal ($\varphi$) and poloidal ($\theta$) directions. The gray boxes, when present, indicate the positions of the UL components (constituted by $8$ units). \textbf{c)} Energy per surface unit carried by the markers onto the plasma-facing components in the $\varphi$-$\theta$ plane. \textbf{d)} Energy per surface unit carried by the markers onto the plasma-facing components  watching from a virtual camera. The camera is placed on the bottom of the tokamak, looking upward, towards the upper part of the tokamak where the ULs are present.}
    \label{fig:Markers1e6}
\end{figure}
\\
We discuss now the results obtained in simulations with the reference number of markers $N_{markers}=10^6$. In \cref{fig:Markers1e6} the RE deposition is shown at the end of the time evolution of the markers during the termination phase. Each row shows the results corresponding to three different simulations with three different plasma-wall interfaces.\\ 
The first line of \cref{fig:Markers1e6} shows the results for a simulation with only AFW and no ULs. In \cref{fig:Markers1e6}~(a) the markers lost to the wall (green points) and those still confined at the end of the termination phase (pink points), are shown in the upper part of the poloidal cross section of the machine. In \cref{fig:Markers1e6}~(b) all markers lost to the wall are shown at the toroidal $\varphi$ and poloidal $\theta$ positions at which they have hit the wall during the simulation. In particular, $\theta$ is calculated with respect to the position of the magnetic axis, as shown by the grey dashed lines in \cref{fig:Markers1e6}~(a). In \cref{fig:Markers1e6}~(c) the energy per surface unit deposited onto the wall by the markers is shown, still in the $\varphi$-$\theta$ plane. In \cref{fig:Markers1e6}~(d) the energy per surface unit deposited by the markers onto the wall is shown by observing from a virtual camera placed at the bottom of the machine and looking towards the upper part of the tokamak.\\ 
The second row of \cref{fig:Markers1e6} shows the results of a simulation where the ULs have also been taken into account. There, we have considered 8 ULs each having a toroidal width of about $\Delta\varphi \approx 11.25^\circ$ and the positions in the R-Z plane  as designed by the DEMO team \cite{RICHIUSA2024114329}, with a protrusion from the AFW of $70$ mm. The positions of the limiters in the $\varphi$-$\theta$ plane are represented by the grey boxes in \cref{fig:Markers1e6}. As the reader can see, with the current plasma configuration the majority of REs do not hit the ULs.
\\ 
It should be stressed that the design of an EU-DEMO is a work in progress. In particular, the simulations presented here have been obtained from an initial plasma scenario for the SOF corresponding to that presented in Ref.~\cite{SICCINIO2022113047}, as already anticipated in \cref{sec:scenario}. The plasma scenario for the SOF was later modified, as already described in Ref.~\cite{MAVIGLIA2022113067,MAVIGLIA2020111713}. Furthermore, the plasma scenario considered in the present work is different from the one that led to the positioning of the ULs, as nicely described in Ref.~\cite{RICHIUSA2024114329}. This explains why, in the present work, the REs do not hit the ULs at the expected positions. Given the difficulty, in terms of numerical and time cost, of repeating these simulations by modifying the initial plasma scenario, in order to estimate the RE energy deposition onto the ULs and to test their effectiveness in shielding the AFW, we have chosen here \q{ad hoc} to move the ULs to the positions in the R-Z plane where, in the present scenario, the majority of the REs hit the wall. In the future, all the simulations performed here will be repeated by changing the initial plasma scenario, in particular by selecting the one that has led to the UL positioning \cite{RICHIUSA2024114329}. In \cref{fig:Markers1e6} we have marked with \q{Current ULs} the results obtained by choosing the UL configurations foreseen by the DEMO team. With \q{Moved ULs} we refer to the results obtained by artificially moving the ULs to the poloidal positions where the majority of REs are lost. We emphasise once again that this is purely an exercise to assess the effectiveness of the ULs in protecting the wall in the present situation. These studies are not intended to suggest that the ULs should be moved to a different position in the poloidal plane. Because other problems would arise, such as the impossibility of integrating and remotely maintaining sacrificial limiters due to the limited space, or the lack of maintenance ports when moving from 1 o'clock to 11 o'clock in the poloidal plane.
\\
Both the sets of plots in \cref{fig:Markers1e6}, where the presence of the ULs has been taken into account, show that the ULs are able to partially or more extensively shield the AFW from the RE energy deposition. The REs remain concentrated to a smaller area, creating spots of accumulated energy deposition. These spots are visible on the right side of the UL blocks when looking at \cref{fig:Markers1e6}~(d) anticlockwise and still on the right side of the boxes in \cref{fig:Markers1e6}~(c). 
\begin{figure}[!h]
    \centering  
    \includegraphics[width=500pt]{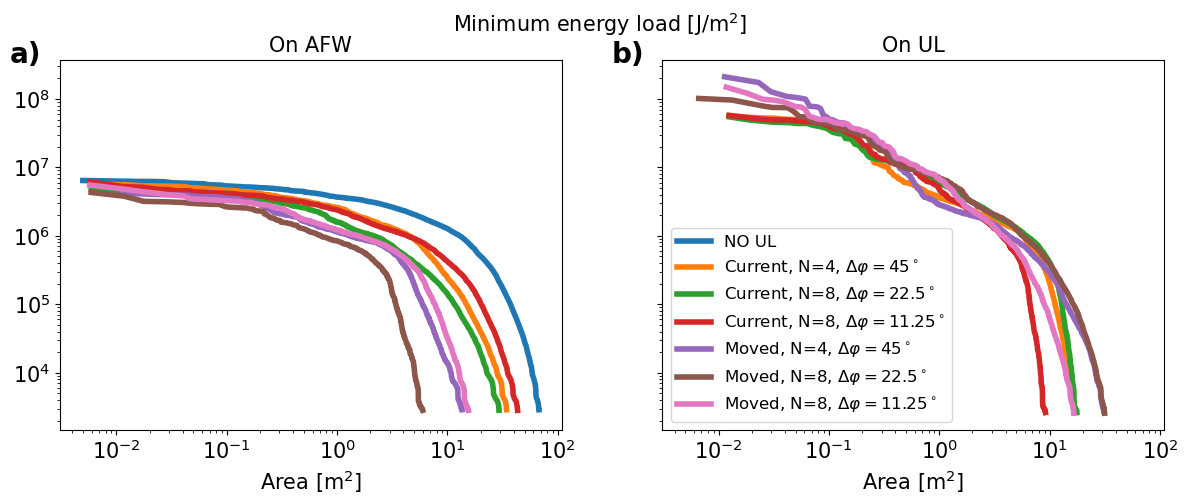}
    \caption{Energy per surface unit that arrives onto a certain area. \textbf{a)} Energy per surface unit onto the AFW only. \textbf{b)} Energy per surface unit onto the ULs only. Note that the blue curve corresponds to the configuration without ULs. In the legend, the different UL configurations considered are shown. \q{Current} refers to configurations with the UL positions in the R-Z plane designed by the DEMO team. \q{Moved} labels the configurations with the ULs shifted \q{ad-hoc} to the position in the R-Z plane where, in the present paper, the majority of REs hit the wall. \q{N} indicates the number of units that constitute the ULs, while $\Delta\varphi$ indicates the width in the toroidal direction of every unit. We remind here that the reference values (as proposed by the DEMO team) are: $N=8$ and $\Delta\varphi=11.25^\circ$.}
    \label{fig:Markers1e6EnLoad}
\end{figure}
\\
In \cref{fig:Markers1e6EnLoad} the minimum energy per surface unit that arrives onto a given surface area is shown for simulations with different plasma-wall interfaces. We have considered configurations with the ULs at the positions in the R-Z plane designed by the DEMO team (labelled \q{Current}). We have also moved the ULs in the R-Z plane to the positions where, in this work, the majority of the REs hit the plasma-facing components (labelled \q{Moved}). We have also varied the number of units $N$ of ULs and their toroidal width $\Delta\varphi$. We remind here that, as designed by the DEMO team, the ULs should be constituted by 8 pieces, each with toroidal width $\Delta\varphi\approx11.25^{\circ}$. This configuration was based on studies of the energy loads due to charged particles during the TQ, whereas the RE loads are estimated for the first time in this paper.
In \cref{fig:Markers1e6EnLoad}~(a) the energy deposition on the AFW only is shown, while \cref{fig:Markers1e6EnLoad}~(b) shows the deposition on the ULs only. Note that the blue curve in \cref{fig:Markers1e6EnLoad} corresponds to the configuration where only the AFW is present and no ULs, cf. the first line in \cref{fig:Markers1e6}. The red and pink curves refer, respectively to the cases with ULs shown in \cref{fig:Markers1e6}. We can see from \cref{fig:Markers1e6EnLoad} that the effect of the ULs is to reduce the AFW area affected by the RE energy deposition. There is also a slight reduction in the maximum energy per surface unit deposited onto the AFW. On the other hand, the maximum energy deposited on the ULs is greatly increased with respect to the configuration without ULs.  This is because, as expected, the area of intersection of the REs with the ULs is smaller and the REs are more concentrated into a smaller area. In \cref{fig:Markers1e6MaxVal} we have summarised some of the results contained in \cref{fig:Markers1e6EnLoad}, showing their dependence on the chosen plasma-wall configuration.
\begin{figure}[!h]
    \centering  
    \includegraphics[width=480pt]{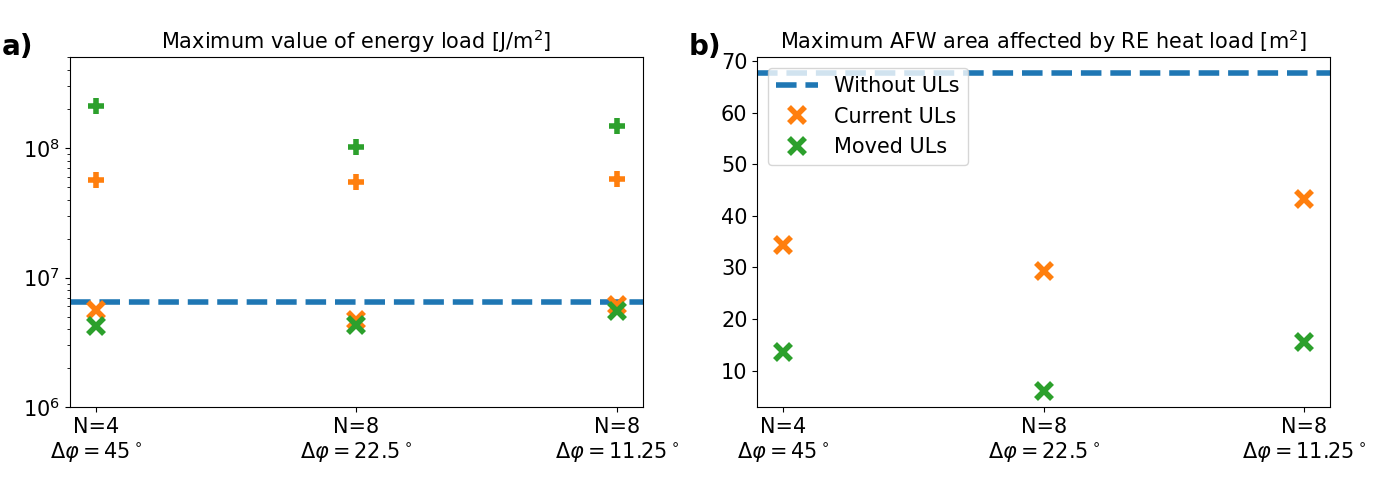}
\caption{Dependence of two meaningful quantities contained in \cref{fig:Markers1e6EnLoad} against the chosen plasma-wall configuration. \textbf{a)} Maximum value of energy load onto AFW in presence of ULs (\q{x}) and on ULs (\q{+}). The blue dashed line represents the maximum value of energy load on the AFW in absence of ULs.  \textbf{b)} Maximum AFW area affected by the REs.}
    \label{fig:Markers1e6MaxVal}
\end{figure}
\\
Finally, \cref{fig:ScanEload}~(a) shows the dependence of the minimum energy load on the chosen number of initialised markers for a simulation with $n\in[0;7]$. In \cref{fig:ScanEload}~(b) the variation of the minimum energy loads in simulations where the reference number of markers $N_{markers}=10^6$ has been kept, but the number of toroidal harmonics kept in the simulations has been varied, is shown.
\newpage

\newpage
\section{Conclusion}\label{sec:conclusion}
In this paper we have carried out numerical simulations using the JOREK code to investigate the effects of REs onto the DEMO plasma-facing components. In particular, we have established the work-flow to assess the damage created by the REs onto the wall, in the presence of ULs shielding this last. We believe the established rather complete workflow to be helpful in the process of validating the ULs. To this end, we first studied, in axisymmetric MHD simulations, the formation of a RE beam in the CQ of a mitigated hot VDE. We allowed the RE beam to evolve until the plasma edge safety factor reached a value close to $2$. At this point, the simulation was restarted with higher toroidal harmonics, in particular keeping the toroidal mode numbers in the range $n\in[0;7]$. In these non-axisymmetric simulations, the decay of the RE beam was observed due to the growth of MHD instabilities that resulted in the stochastization of the magnetic field lines. Later, the reformation of the RE beam was also observed. In post-processing, particle tracing has been used to evolve the initialised RE markers during the RE beam termination phase and to deposit their thermal loads onto the plasma-facing components in case of collisions with the latter. We have considered the current plasma wall configuration with the ULs in the position designed by the DEMO team, but we have also considered other configurations with different number of UL units and unit toroidal extension.
 In the presented studies, we have not taken into account the effects of the RE regeneration (observed in our simulations) after the first loss event and the associated conversion of magnetic energy into kinetic RE energy that would be also then deposited to the wall.
Our studies clearly show that the ULs are important to protect the FW from the REs, shielding it from most of them, reducing the FW area affected by REs.  Nevertheless, while the maximum energy per surface unit onto the FW is slightly reduced, the maximum energy per surface unit onto the ULs is almost $1$ order of magnitude higher than that onto the FW in absence of ULs. In our studies we do not consider a detailed CAD model for FW and ULs and, as shown in Ref.~\cite{Hannes2024} for ITER studies, when considering the details of the plasma-facing components, the RE localisation is further increased, leading to a consequent increase in the deposited energy per surface unit.
Further studies will be carried out to support the development of the design of the ULs to adequately shield the FW.

\section{Acknowledgements}

The authors are grateful to Dr. R. Ramasamy, Dr. A. Cathey and V. Mitterauer for useful and interesting discussions.\\
This work has been carried out within the framework
of the EUROfusion Consortium, funded by the
European Union via the Euratom Research and
Training Programme (Grant Agreement No 101052200
— EUROfusion). Views and opinions expressed are
however those of the author(s) only and do not
necessarily reflect those of the European Union or the
European Commission. Neither the European Union
nor the European Commission can be held responsible
for them.\\ 
Many of the simulations were done on the
Marconi-Fusion supercomputer hosted at CINECA.
\clearpage

\appendix
\section{Appendix}\label{sec:appendix}
The (normalized) equations taken into account in this work, governing the evolution of the reduced single-fluid MHD plasma model coupled to the impurities and RE equations, are:
\begin{equation}
    \frac{1}{R}\partial_t\psi = \frac{\eta}{R}\left(j-\frac{c\,F_{0}}{B\,R}\,n_{RE}\right) + \left[\psi,u\right]-\frac{F_0}{R}\,\partial_{\varphi}u-\partial_R(\eta_{num}\partial_R j)-\partial_Z(\eta_{num}\partial_Z j)
\label{Eq1}
\end{equation}

\begin{align}
R\nabla\cdot\left(R^2\rho\partial_t\nabla_{pol}\partial_t u\right)=&
    \left[\frac{R^2}{2}|\nabla_{pol}u|^2,\rho R^2\right]+\left[\rho R^4\omega,u\right]+\left[\psi,j\right]-\frac{F_0}{R}\partial_{\varphi}j
    -\left[R^2,\rho T\right]+R\nabla\cdot\left(\mu\nabla_{pol}\omega\right)
    \label{Eq2}
\end{align}

\begin{equation}
    j=\Delta^* \psi\equiv R^2\nabla\cdot\left(\frac{1}{R^2}\nabla_{pol}\psi\right)
    \label{Eq3}
\end{equation}

\begin{equation}
    \omega=\Delta_{pol}u\equiv \nabla\cdot \nabla_{pol}u
    \label{Eq4}
\end{equation}

\begin{align}
    \partial_t\rho=S_\rho + S_{imp}+ R\left[\rho,u\right]+2 \rho\partial_Z u &+ \nabla\cdot\left[D_{\parallel}\nabla_{\parallel}(\rho-\rho_{imp})+D_\perp\nabla_{\perp}(\rho-\rho_{imp})\right]
    \label{Eq5}
    \\
    \nonumber
    &+ \nabla\cdot\left[D_{\parallel,imp}\nabla_{\parallel}\rho_{imp}+D_{\perp,imp}\nabla_{\perp}\rho_{imp}\right]
\end{align}

\begin{align}
    \partial_t p =& R [\rho T,u] +2\gamma\rho T\,\partial_Z u +\nabla\cdot\left(\chi_{\parallel}\nabla_{\parallel}T+\chi_{\perp}\nabla_{\perp}T\right)
    +(\gamma-1)\frac{\eta_{Ohm}}{R^2}\left(j-\frac{c\,F_0}{B\,R}n_{RE}\right)^2
    \label{Eq6}
    \\
    \nonumber
    &+(\gamma-1) \Biggl\{E_{ion}\left(2\rho_{imp}\,\partial_{Z}u + R\,[\rho_{imp},u]\right)+\nabla\cdot\left(E_{ion}\,D_{\parallel,imp}\,\nabla_{\parallel}\rho_{imp}+E_{ion}\, D_{\perp,imp}\,\nabla_{\perp}\rho_{imp}\right)\Biggr\}
    \\
    \nonumber
    &+(\gamma-1) \Biggl\{E_{ion}^{bg}\left(2(\rho-\rho_{imp})\,\partial_{Z}u + R\,[\rho-\rho_{imp},u]\right)+\nabla\cdot\left(E_{ion}^{bg}\,D_{\parallel}\,\nabla_{\parallel}(\rho-\rho_{imp})+E_{ion}^{bg}\, D_{\perp}\,\nabla_{\perp}(\rho-\rho_{imp})\right)\Biggr\}
    \nonumber
    \\
    \nonumber
     &+ \alpha_{imp,bis}\rho_{imp}\,R\,[T,u]+\alpha_{imp}\,T\,R\,[\rho_{imp},u]+2\gamma\,\alpha_{imp}\rho_{imp}T\partial_Z u+\frac{\gamma-1}{2}R|\nabla_{pol}u|^2\left(S_{bg}+S_{imp}\right)\\
     \nonumber
    &-(\rho+\beta_{imp}\rho_{imp})(\rho-\rho_{imp})L_{rad,Dcont}
    -(\rho+\beta_{imp}\rho_{imp})\left[f_{rad,bg}+\rho_{imp}L_{rad}\right]+(\gamma-1)R\,\rho_{imp}\frac{dE_{ion}}{dT}[T,u]
    \nonumber
\end{align}
\begin{equation}
\partial_{t}\rho_{imp}=\nabla\cdot\left[D_{\parallel,imp}\nabla_{\parallel}\rho_{imp}+D_{\perp,imp}\nabla_{\perp}\rho_{imp}\right]+R[\rho_{imp},u]+2\rho_{imp}\partial_{Z}u+S_{imp}
\label{Eq7}
\end{equation}
\begin{align}
    \partial_t n_{RE}= S_{RE}+S_{avalanche} + 2\,n_{RE}\partial_Z u + R\left[n_{RE},u\right]
 +\nabla\cdot\left(D_{\parallel,RE}\nabla_{\parallel}n_{RE}+D_{\perp,RE}\nabla_{\perp}n_{RE}\right)\quad .
\label{Eq8}
\end{align}
In \cref{Eq1,Eq2,Eq3,Eq4,Eq5,Eq6,Eq7,Eq8} the Poisson bracket and the gradient in the R-Z plane have been introduced:
\begin{equation}
    [f,g]=\hat{\boldsymbol{e}}_\varphi\cdot\nabla f\times\nabla g=\partial_R f\partial_Z g - \partial_R g\partial_Z f,\quad \nabla_{pol}\,h=\hat{\boldsymbol{e}}_R\partial_R h + \hat{\boldsymbol{e}}_Z\partial_Z h\,.
\end{equation}
$\eta_{num}$ is the hyperresistivity, $L_{rad}$ and $L_{rad,Dcont}$ are impurity and Deuterium ions radiation respectively, while $E_{ion}$ and $E_{ion,bg}$ are the ionization energies of impurities and background ions respectively. The other important parameters and their values used in the simulations presented in this work are given in \cref{Table_modifications}. Finally: 
\begin{equation}
    \alpha_{imp}=\frac{m_i}{2\,m_{imp}}(\langle Z_{imp}\rangle+1)-1,\quad \alpha_{imp,bis}=\alpha_{imp}+T\frac{d}{dT}\alpha_{imp},\quad \beta_{imp}=\frac{m_i}{m_{imp}}\langle Z_{imp}\rangle-1\,,
\end{equation}
being $\langle Z_{imp}\rangle$ the average impurity charge, $m_i$ and $m_{imp}$ the ion and impurities  masses respectively. In the present paper, we consider a deuterium ion plasma.\\
In \cref{Eq8} we do not model the RE parallel transport as an advection at the speed of light but rather as a parallel diffusion. By means of this choice, we are able to reduce the computational cost associated with the modelling of RE parallel transport as explained in Ref. \cite{PhysRevVinodh}.

\begin{table}[!h]
\begin{center}
\caption{Parameters in use before and after the ATQ.}
\label{Table_modifications}
\begin{tabular}{||c c c c ||} 
 \hline
   Parameter & Dependency & Value & Description \\ [0.5ex] 
 \hline\hline
 $D$ & Constant & $1.04$ m$^2$ s$^{-1}$ & \makecell{(Isotropic) particle diffusion coefficients \\ for thermal plasma and impurities} \\ [1ex] 
 \hline
 $D_{\parallel,RE}$ & Constant & $1.54\times 10^9$ m$^2$ s$^{-1}$ &Parallel RE  diffusion coefficient \\ [1ex] 
 \hline
 $D_{\perp,RE}$ & Constant & $1.54\times 10^{-2}$ m$^2$ s$^{-1}$ & Perpendicular RE diffusion coefficient\\ [1ex] 
 \hline
 $\chi_\parallel$ & Spitzer-Haerm $\propto T_e^{5/2}$ & \makecell{$\chi_\parallel^{max}=\chi_{\parallel}(876\,eV)$ \\ For $T_e>876\,eV,\,\chi_{\parallel}=\chi_{\parallel}^{max}$} & Parallel heat diffusion coefficient \\ [1ex] 
 \hline
 $\chi_\perp$ & Profile & $\chi_{\perp,core}=0.5$ m$^{2}$ s$^{-1}$ & Perpendicular heat diffusion coefficient \\ [1ex] 
 \hline
 $\eta$ & Spitzer $\propto T_e^{-3/2}$ & \makecell{$\eta^{min}=\eta(1.26\,keV)$ \\ For $T_e>1.26\,keV,\,\eta=\eta^{min}$} & Plasma resistivity  \\ [1ex] 
 \hline
 $\mu$ & Constant & $\mu_{core}=2.47\times 10^{-3}$ kg m$^{-1}$ s$^{-1}$ & Viscosity \\ [1ex] 
 \hline
\end{tabular}
\end{center}
\end{table}




 
\newpage
\clearpage
\bibliographystyle{unsrturl}
\bibliography{literature}
\end{document}